\documentclass{ws-procs9x6}
\def\delslash{\,\,{\raise.15ex\hbox{/}\mkern-10mu \partial}}

\begin{document}

\title{Universality in fermionic field theries at finite temperature}

\author{Costas G. STROUTHOS\footnote{\uppercase{S}upported by 
\uppercase{L}everhulme \uppercase{T}rust}} 

\address{Department of Physics, 
University of Wales Swansea, \\ 
Singleton Park, Swansea SA2 8PP, U.K.}


\maketitle

\abstracts{
We discuss the critical properties of the three-dimensional NJL model
at nonzero temperature.
We show that the $Z_2$-symmetric model undergoes a second order phase 
transition 
with $2d$ Ising exponents and its critical region is suppressed by a 
factor $1/\sqrt{N_f}$.
We also provide numerical evidence that the $U(1)$-symmetric model 
undergoes a BKT transition 
in accordance with the dimensional reduction scenario.
}


Phase transitions in QCD at nonzero temperature have been studied
intensively over the last decade.
However, since the problem of chiral symmetry breaking and its restoration 
is intrinsically non-perturbative, the number of available techniques is 
limited and most of our knowledge about this phenomenon comes from lattice
simulations. 

Although there is little disagreement that the chiral phase transition
in QCD with two massless quarks is second order, no quantitative
work or simulations have been done that decisively determine its universality
class. Universality arguments are very appealing due to their beauty
and simplicity and they can be phrased as follows: At finite
$T$ phase transitions, the correlation length diverges in the transition
region and the long range behavior is that of the $(d-1)$ classical spin model
with the same symmetry, because the IR region of the system is
dominated by the zero mode of the bosonic field. The contribution
of non-zero modes does not affect the critical singularities but can be
absorbed into non-universal aspects of the transition.
Consequently, fermions which satisfy anti-periodic boundary conditions
and do not have zero modes are expected to decouple from the scalar sector.
A possible loophole to this standard scenario is that
the mesons are composite $f \bar{f}$ states and their size and density
increase as $T \rightarrow T_c$.
Therefore, if the transition region can be described as a system of
highly overlapping composites
the violation of the bosonic character of the mesons may be maximal
and the fermions become essential degrees of freedom irrespective of
how heavy they are.

The NJL model has been proved to be an interesting and tractable
laboratory to study phase transitions both numerically by means of lattice
simulations and analytically in the form of large-$N_f$ expansions \cite{rosen91,hands93}.
The lagrangian density of the $U(1)$-symmetric model is
\begin{equation}
\mathcal{L}= \bar{\psi}_i(\partial\hskip -.5em / + 
\sigma + i \gamma_5 \pi)
\psi_i
+ \frac{N_f}{2g^2} (\sigma^{2}+ \pi^2),
\label{eq:L}
\end{equation}
where the index $i$ runs over $N_f$ fermion flavors. There are several 
features which make this model interesting for the modelling of strong 
interactions: (i) The spectrum of excitations contains both ``baryons''
and mesons, namely the elementary fermions $f$ and the composite $\bar{f}f$
states; (ii) for sufficiently strong coupling $g^2>g_c^2$ it exhibits spontaneous
chiral symmetry breaking implying dynamical generation of a fermion mass; 
(iii) for $2<d<4$ there is an interacting continuum limit at a critical  
coupling, which for $d=3$ has a numerical value $g^2_c \approx 1.0/a$ in the large-$N_f$
limit if a lattice regularization is employed.
At leading order in $1/N_f$ the model  
undergoes a second order transition
at $T_c=\frac{M_f}{2 \ln{2}}$ \cite{klimenko} ($M_f$ is the dynamical fermion mass 
at $T=0$) with Landau-Ginzburg mean field
scaling. This result is valid for both continuum and discrete chiral symmetries.

We studied the critical behavior of the  $Z_2$-symmetric model with finite $N_f$
by performing lattice simulations near $T_c$ \cite{stephanov}. 
We simulated the model exactly for $N_f=12$
with the Hybrid Monte Carlo method. The staggered fermion lattice action
and further details concerning the algorithm can be found in \cite{hands93}.
The temporal lattice size was $L_t=6$ and the spatial size varied from $L_s=18$ to $50$.
The expectation value of the auxiliary sigma field 
 serves as a convenient order parameter for the
theory's critical point.
By using finite size scaling (FSS) techniques we extracted the critical exponents.
The results
which are summarized in Table 1 support the dimensional reduction scenario,
because the values of the exponents are in good
agreement with those of the $2d$ Ising model rather than the mean field
theory ones. This implies that the role of composite mesons is relevant near $T_c$ even when 
$N_f$ is large but finite. 
Simulations with the meron-cluster algorithm of the $4d$ NJL model gave 
$3d$ Ising exponents with very high accuracy.  
\cite{cluster1}.

\begin{table}[t]
\label{tab:1}       
\setlength{\tabcolsep}{0.8pc} 
\begin{tabular}{|l|lll|}
\hline
Exponents & FSS & $Z_2$ & MF \\ 
\hline
$\nu$     & 1.00(3) & 1 & 0.5 \\
$\beta_m/\nu$ & 0.12(6) & 0.125 & 1 \\
$\gamma/\nu$ & 1.66(9) & 1.75 & 2 \\
\hline
\end{tabular}
\end{table}

Next we tried to understand how the large-$N_f$ mean field theory
prediction reconciles with the dimensional reduction and universality arguments.
The answer is that the large-$N_f$
description has its applicability region. We studied this in detail for a Yukawa theory
\cite{stephanov} and we showed that
the phenomenon which leads to an apparent contradiction is the suppression of the
width of the non-mean field critical region by a power of $1/N_f$.
The $2d$ Ising critical behavior sets in when $T \gg M_{\sigma}(T)$ ($M_{\sigma}(T)$ is the
thermal mass of the $\sigma$ meson).
If the cutoff $\Lambda \gg T$ the renormalized self-interaction coupling $\lambda(T)$
in the large-$N_f$ limit is close to the IR fixed point of the Yukawa theory
and is given by
$\lambda(T) \sim T^{4-d}/N_f$ for $2\!<\!d\!<\!4$.
The mean field approximation breaks down because of self-inconsistency
when the value of the coupling of the $(d-1)$ dimensional scalar theory
$\lambda_{d-1} \sim T \lambda(T)$ on the scale $M_{\sigma}^{5-d}(T)$
(the power $d-5$ comes from
dimensional analysis) is not small anymore. Therefore,
for $d=3$ the Ginzburg criterion for the applicability
of the mean field scaling is given by $M_{\sigma}(T) \gg T/\sqrt{N_f}$.
This scenario was also verified numerically by measuring 
$\langle \sigma \rangle = M_f(T) = M_{\sigma}(T)/2$
at the crossover into the mean field region \cite{stephanov}.

Additional evidence in favor of the dimensional reduction scenario
was produced in studies of the $U(1)$-symmetric NJL$_3$ model \cite{babaev,strouthos01,yamamoto}.
Both analytical and numerical  results showed that
its phase structure at $T \neq 0$ is the same as that of the $2d$ $XY$ model.
As we already mentioned, at leading order in $1/N_f$ this  model undergoes 
a second order chiral phase
transition \cite{rosen91}.
This conclusion, however, is expected to be valid only when $N_f$ is strictly infinite, i.e. 
when the fluctuations of the
bosonic fields are neglected, otherwise it runs foul of
the Coleman-Mermin-Wagner (CMW) theorem, which states that in two-dimensional systems
the continuous
chiral symmetry must be manifest for all $T \! > \! 0$.
Next-to-leading order calculations \cite{babaev,yamamoto} demonstrated that
the model undergoes a Berezinskii-Kosterlitz-Thouless (BKT) transition \cite{kosterlitz},
in accordance with the dimensional reduction scenario.
It is easy to visualize the BKT scenario if we use the ``modulus-phase'' parametrization:
$\sigma + i \pi \equiv \rho e^{i \theta}$.
In two spatial dimensions logarithmically divergent infrared fluctuations do not allow the phase
$\theta$ to take a fixed direction and therefore prevent spontaneous symmetry
breaking via $\langle\theta\rangle\not=0$.
The critical temperature $T_{BKT}$ is expected to separate two \emph{different} 
chirally symmetric phases:
a low $T$ phase, which is characterized by power law phase correlations
$\langle e^{i\theta(x)} e^{-i\theta(0)} \rangle \sim x^{-\eta(T)}$ at distances $x \gg 1/T$
but no long range order (i.e.
a spinwave phase where chiral symmetry is ``almost but not quite broken''), and a high $T$
phase which is characterized by exponentially decaying phase correlations with no long range order.  In
other words for $T \leq T_{BKT}$
there is a line of critical points characterised by 
$\eta(T) = \frac{f(T)}{N_f}$ \cite{appelquist} for $T \ll T_{\rm BKT}$ and at $T=T_{\rm BKT}$ 
$\eta(T)$ is expected to be $0.25$.

\begin{figure}[t]
\begin{tabular}{ll}
{\includegraphics[scale=0.29]{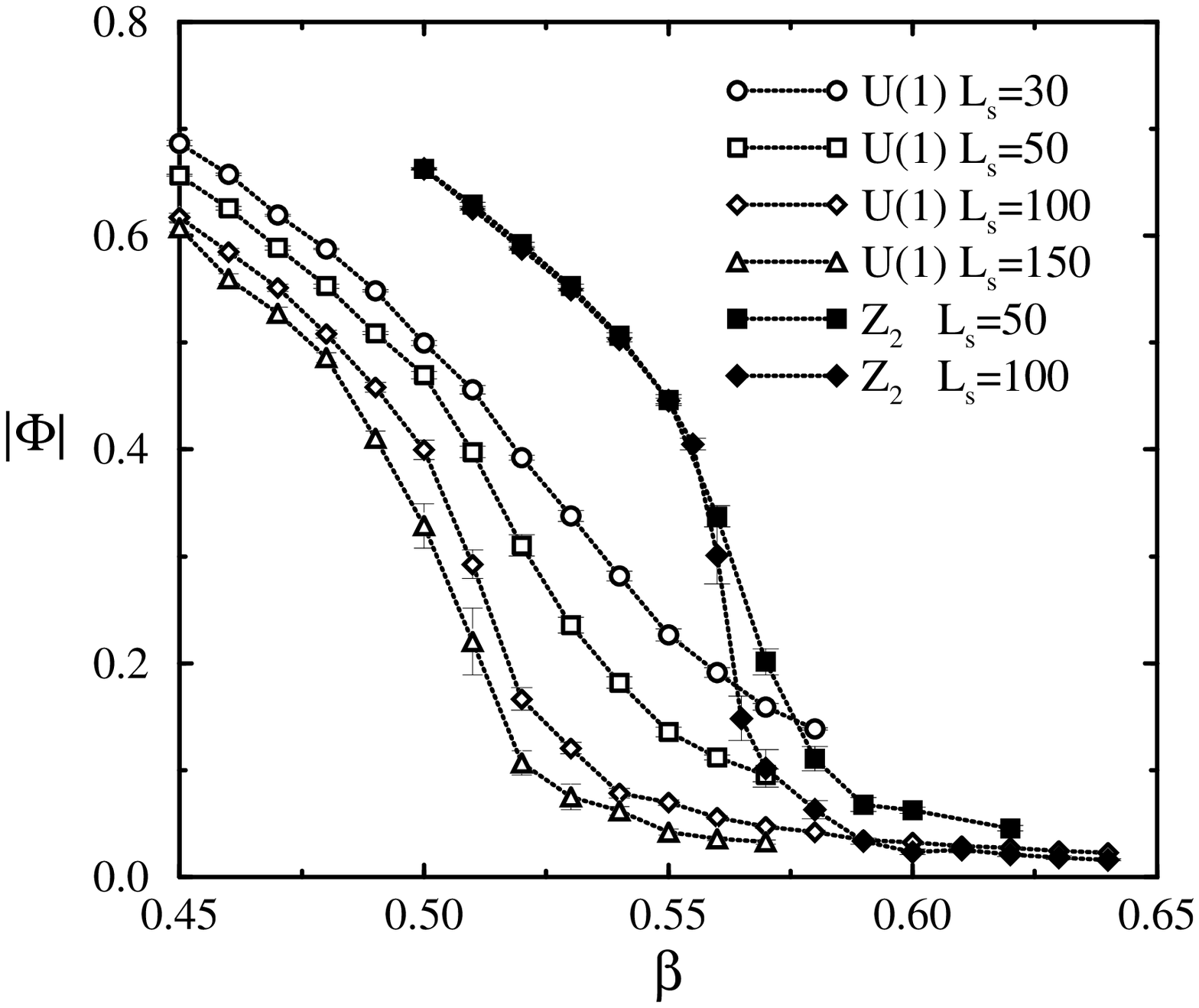}}&
{\includegraphics[scale=0.29]{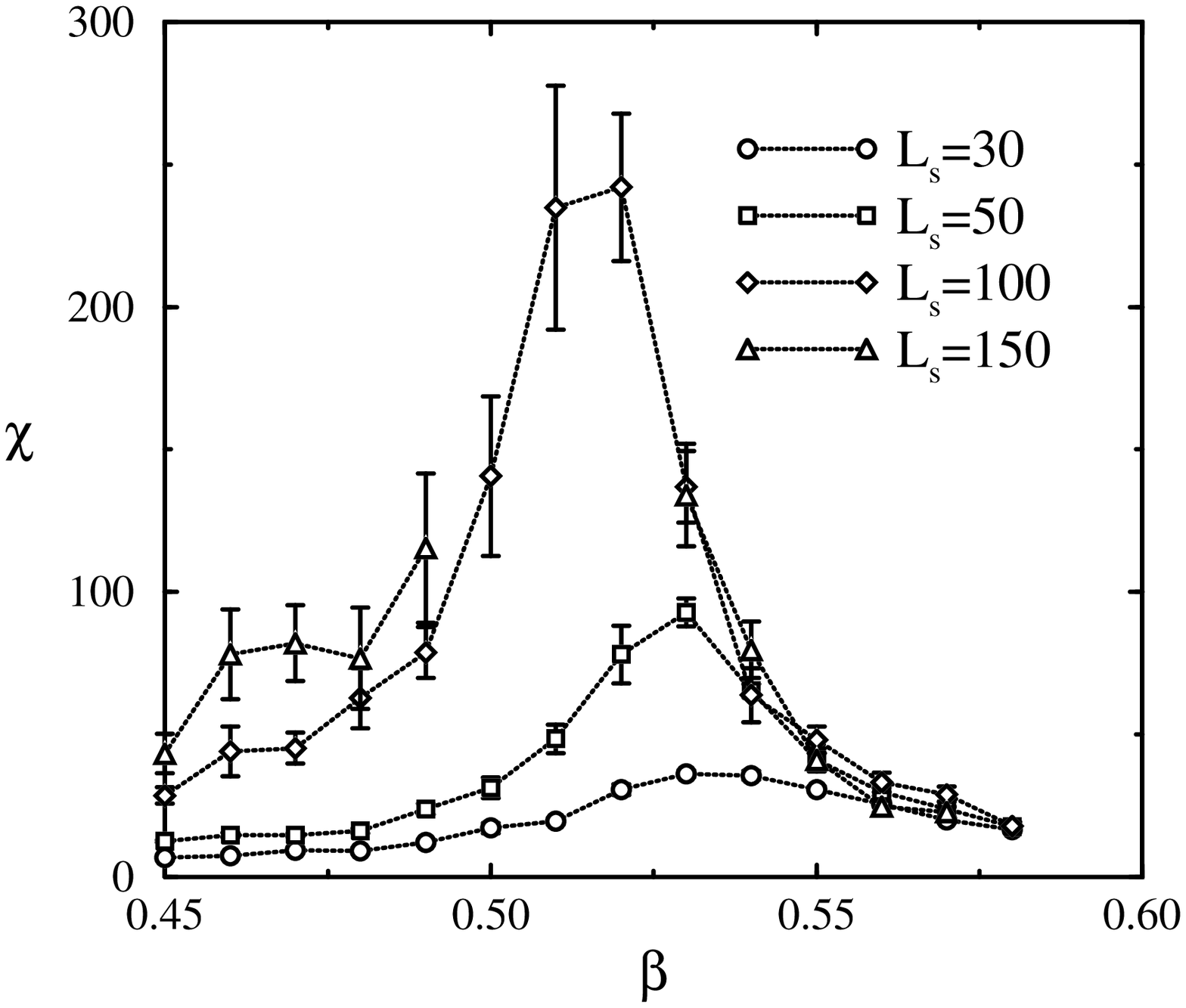}}
\end{tabular}
\caption{Order parameter and susceptibility vs. coupling $\beta$ for various values of $L_s$.}
\label{fig:results}
\end{figure}

In order to study the behavior of the chiral symmetry
at $T\!>\!0$ in the absence of a fermion bare mass in the Lagrangian
the best thing to measure is
an effective ``order parameter'' $|\Phi| \equiv \sqrt{\sigma^2 + \pi^2}$,
which is a projection onto
the direction of $\Phi^{\alpha}  \equiv (\sigma, \pi)$ separately for
each configuration.
In Fig.\ref{fig:results} we plot $|\Phi|$ versus $\beta \! \equiv \! 1/g^2$
for fixed lattice temporal size $L_t=4$ and different lattice spatial sizes 
together with the results from simulations
of the model with a $Z_2$ chiral symmetry. In both cases $N_f=4$.
At the bulk critical coupling
$\beta_c^{\rm bulk}\! \equiv 1/g^2 \! \approx \! 0.86$ \cite{strouthos01}
the lattice spacing becomes zero and $T \! \rightarrow \! \infty$.
It is clear that
the order parameter of the $Z_2$ model is independent of the lattice size until just before the
transition at $\beta_c^{Z_2} \!= \! 0.565(3)$,
whereas in the $U(1)$ model $|\Phi|$ has a strong size dependence for a large range of values
of $\beta$, i.e. it
decreases rapidly as the spatial volume increases in accordance with the expectation that chiral
symmetry should be restored for $T\!>\!0$.
The finite spatial extent $L_s$ provides a cut-off for the divergent correlation length and according to
the BKT scenario
the slow decay of the correlation function $\langle e^{i\theta(x)} e^{-i\theta(0)} \rangle$
for $T\!<\!T_{BKT}$ ensures a non-zero magnetization
even in a system with very large size.
In Fig.\ref{fig:results} we plot the susceptibility of the order parameter
$\chi = V (\langle |\Phi|^2 \rangle - \langle |\Phi| \rangle^2)$, measured on
lattices with different spatial size. We observe from this figure that:
(a) there is a phase transition since the peak of $\chi$ diverges
as we increase $L_s$; 
and (b) in the low $T$ phase the susceptibility has a stronger size dependence
than in the high $T$ phase and the errors at low $T$ are much larger than
at high $T$. This is strong evidence that the system is critical
in the low $T$ phase in accordance with the BKT scenario.
It is instructive to compare the results for the $U(1)$ model 
with results extracted from recent simulations of the 
$SU(2) \times SU(2)$-symmetric NJL$_3$ model \cite{walters}. In the latter case
the correlation length is finite for $T>0$ 
and the effective order parameter goes to zero more rapidly 
as we increase $L_s$ than in the $U(1)$ case. 
It was also shown that both models have a pseudogap phase where the fermions  
remain massive in the absence of long range correlations \cite{babaev,strouthos01,walters}.
Recently, simulations of an extension of the attractive Hubbard model with the 
meron-cluster algorithm confirmed with high accuracy that 
$0 \leq \eta \leq 0.25$ \cite{cluster2}. 

In summary, we have shown that the dimensional reduction scenario describes correctly the
phase transition of simple fermionic field theories with composite mesons. We plan to extend our
work to the four-dimensional NJL model and study the dependence of the width  
of the critical region on the degree of chiral symmetry. It has already been shown that 
in the $(3+1)d$ $O(N)$ sigma model the critical region is much wider for $N \geq 2$ than for
the discrete symmetry case \cite{tziligakis}. Finally, it will be interesting to understand 
the issue of universality in two-flavor QCD.  
Simulations with massless quarks of QCD with a four-fermi interaction \cite{sinclair}
near the continuum limit may provide decisive results in this direction.

\end{document}